\documentstyle[12pt,frascatiphys]{article}
\begin{document}
\title{ 
RADIAL EXCITATIONS
}
\author{
T. Barnes \\
{\em Physics Division, Oak Ridge National Laboratory} \\ 
{\em Oak Ridge, TN 37831-6373, USA } \\ 
{\em Department of Physics , University of Tennessee  } \\ 
{\em Knoxville, TN 37996-1501, USA} \\ 
}
\maketitle
\baselineskip=14.5pt
\begin{abstract}
In this contribution I discuss
recent experimental developments in the spectroscopy of higher-mass
mesons, especially 
candidate radial excitations discussed at the WHS99 meeting
in Frascati.
\end{abstract}
\baselineskip=17pt
\section{Introduction: Why radials?}
We now have strong evidence for a true $J^{PC}=1^{-+}$ exotic
at 1.6 GeV\cite{Chung,Dorofeev,hybrid_pi1} in $\rho\pi$ at BNL and VES, 
and $\eta'\pi$ and $b_1\pi$ at VES, and with a 
possible lighter
state at 1.4 GeV in $\eta\pi$ reported by BNL and Crystal 
Barrel\cite{Chung,CB_etapi,Thoma}. Hadron spectroscopy may 
have finally
found the hybrid mesons anticipated by theorists.
Of course there is an unresolved concern that these 
experimental masses are somewhat lighter than theoretical expectations;
both the flux-tube model\cite{hybr_ft_mass} and recent
LGT calculations\cite{hybr_latt}  
find that the lightest
exotic should be a $J^{PC}=1^{-+}$, {\it albeit} with a mass of
$\approx 1.9-2.0$~GeV.
 
Since $q\bar q g$ hybrids span flavor nonets, there will be many more
such states if this is indeed a correct interpretation of the data.
Specific models of hybrids such as the flux-tube model and the bag model
anticipate that there should be hybrid flavor nonets with
all $J^{PC}$, the majority having nonexotic quantum numbers.
In the flux tube model the lightest hybrid multiplet includes the
nonexotic states
\begin{equation}
J^{PC}({\rm flux-tube\ hybrid\ nonexotics\ }) =
0^{-+}, 1^{--}, 1^{++}, 1^{+-}, 2^{-+}
\label{eq1} 
\end{equation}
and in the bag model the lightest hybrid multiplet includes
\begin{equation}
J^{PC}({\rm bag-model\ hybrid\ nonexotics\ })  =
0^{-+}, 1^{--}, 2^{-+}
\ .
\label{eq2} 
\end{equation}

In addition to hybrids we also expect glueball degrees of freedom,
which will overpopulate the I=0 sector relative to expectations for
$q\bar q$.
The spectrum of ``quenched" glueballs
(in the absence of quarks) is reasonably well established from
LGT studies\cite{LGT_glueballs}; 
the lightest state is a scalar at about 1.6 GeV,
followed by a
$0^{-+}$ and a
$2^{++}$ at about 2.2-2.3 GeV. The scalar glueball has been speculatively
identified with the $f_0(1500)$\cite{AC} or alternatively the 
$f_0(1710)$,\cite{SVW} 
and especially in the
$f_0(1500)$ case one needs to invoke important $n\bar n \leftrightarrow G
\leftrightarrow s\bar s$ mixing to explain the 
observed decay branching fractions.

Finally, there is 
a spectrum of weakly bound
molecular states analogous to the $K\bar K$ molecule candidates\cite{Wei90}
$f_0(980)$ and $a_0(980)$, which is at least as extensive as the 
Nuclear Data Sheets.
Unlike glueballs,
the spectrum of molecular states beyond 
$K\bar K$ and the nuclei and hypernuclei 
has received little
theoretical attention. 
There are some quark-model calculations that indicate 
that
vector meson pairs may bind\cite{vecmolec} 
but there has been no systematic investigation of
the expected spectrum.

As a background to these various hadron exotica we have 
a spectrum of conventional $q\bar q$ states, which must be identified
if we are to isolate non-$q\bar q$ exotica. Since the lightest
non-$q\bar q$ states typically have masses and quantum numbers
which are expected for radially excited  $q\bar q$ states, it is
especially important to identify radial excitations.

Identification of the $q\bar q$
and non-$q\bar q$ states in the spectrum will require that we clarify meson
spectroscopy to masses of at least 2.5 GeV, so that the pattern of glueballs,
hybrids and multiquarks can be established through the identification
of sufficient examples of each type of state.

\section{States reported in the WHS99 ``Radial Excitations" session}

\subsection{$p\bar p$ annihilation in flight}

In this session we heard new results from Crystal Barrel $p\bar p$ 
annihilation data in flight, as analyzed by
D.Bugg and collaborators\cite{Bugg}, in several 
final states. The final states discussed
were $\pi\pi$, $\eta\eta$, $\eta\eta'$, $3\pi^o$, $\eta\pi^o$,
$\eta' \pi^o$ and
$\eta \pi^o \pi^o $. 
Some very interesting results were reported, which 
will allow us
to quote some new estimates for the masses of previously unknown higher-mass
$n\bar n$ quarkonium multiplets.

The I=0 states reported by Bugg in 
$\pi^+\pi^-$, $\pi^o\pi^o$, $\eta\eta$, $\eta\eta'$ and
$\eta\pi^o\pi^o$
(taken from a recent preprint\cite{Aniso99}) 
are summarized in 
Table~1.
Bugg also reported results for I=1 states seen in $\pi^+\pi^-$,
$3\pi^o$, $\eta\pi^o$ and $\eta'\pi^o$, given in Table~2.

\begin{table}
\centering
\caption{\it I=0 states reported in Crystal Barrel data in 
$p\bar p\to PsPs$ and $\eta\pi^o\pi^o$ by Bugg et al.
}
\vskip 0.1 in
\begin{tabular}{|l|c|c|c|} \hline
 $J^{PC}$         &  $M$(MeV) & $\Gamma$(MeV) & comments  \\
\hline
\hline
 $6^{++}$   & $ 2530(40)$  &   $250(60)$ & weak $\pi\pi$  \\
\hline
 $4^{++}$   & $ 2335(20)$  &   $150(35)$ &   \\
 $4^{++}$   & $ 2025(15)$  &   $180(15)$ &   \\
\hline
 $3^{++}$   & $ 2280(30)$  &   $210(30)$ & $\eta\pi^o\pi^o$  \\
 $3^{++}$   & $ 2000(40)$  &   $250(40)$ & "   \\
\hline
 $2^{++}$   & $ 2365(30)$  &   $300(50)$ &   \\
 $2^{++}$   & $ 2240(40)$  &   $170(50)$ & $\eta\pi^o\pi^o$   \\
 $2^{++}$   & $ 2210(40)$  &   $310(45)$ &   \\
 $2^{++}$   & $ 2065(30)$  &   $225(30)$ &   \\
 $2^{++}$   & $ 1945(30)$  &   $220(40)$ &   \\
\hline
 $2^{-+}$   & $ 2300(40)$  &   $270(40)$ & $\eta\pi^o\pi^o$   \\
 $2^{-+}$   & $ 2040(40)$  &   $190(40)$ & "   \\
\hline
 $1^{++}$   & $ 2340(40)$  &   $340(40)$ & $\eta\pi^o\pi^o$   \\
\hline
 $0^{++}$   & $ 2335(25)$  &   $225(40)$ & very weak $\pi\pi$  \\
 $0^{++}$   & $ 2095(10)$  &   $190(12)$ &   \\
\hline
\end{tabular}
\label{table1}
\end{table}
\begin{table}
\centering
\caption{ \it I=1 states reported in Crystal Barrel data in 
$p\bar p\to \pi^+ \pi^- $, $3\pi^o $, $\eta\pi^o $  and
$\eta'\pi^o$ by Bugg et al.
}
\vskip 0.1 in
\begin{tabular}{|l|c|c|c|} \hline
 $J^{PC}$         &  $M$(MeV) & $\Gamma$(MeV) & comments  \\
\hline
\hline
 $5^{--}$   & $ 2335(20)$  &   $150(35)$ & $\pi^+\pi^-$  \\
\hline
 $4^{++}$   & $ 2280(20)$  &   $205(25)$ &   \\
 $4^{++}$   & $ 2015(25)$  &   $305(80)$ &   \\
\hline
 $3^{++}$   & $ 2310(40)$  &   $180{+120\atop -60}$ &   \\
 $3^{++}$   & $ 2070(20)$  &   $170(40)$ &   \\
\hline
 $3^{--}$   & $ 2215(35)$  &   $340(55)$ & $\pi^+\pi^-$    \\
 $3^{--}$   & $ 1950(15)$  &   $150(25)$ & "   \\
\hline
 $2^{++}$   & $ 2270(30)$  &   $260(50)$ &   \\
 $2^{++}$   & $ 2080(20)$  &   $235(45)$ &   \\
 $2^{++}$   & $ 1990(30)$  &   $190(50)$ &   \\
\hline
 $1^{++}$   & $ 2100(20)$  &   $300{+30\atop -60}$ &   \\
 $1^{++}$   & $ 2340(40)$  &   $230(70)$ &   \\
\hline
 $1^{--}$   & $ 2270(60)$  &   $260(65)$ & $\pi^+\pi^-$     \\
 $1^{--}$   & $ 2015(45)$  &   $270(115)$ & "   \\
\hline
 $0^{++}$   & $ 2025(30)$  &   $330(75)$ &   \\
\hline
\end{tabular}
\label{table2}
\end{table}

If these results are confirmed, they represent a considerable
contribution to the determination of the
$n\bar n$ quarkonium spectrum
in the mass region of 1.9-2.5 GeV, which is especially relevant to searches 
for glueballs and {\it excited} hybrids.

\subsection{$\tau$ hadronic decays at CLEO: the $a_1(1700)$}

In addition to evidence for radially excited states in $p\bar p$
annihilation, we also heard results from R.Baker\cite{Baker} 
about the possible evidence
for a radial excitation, the $a_1(1700)$, in $\tau$ hadronic decays.
The process discussed was $\tau^- \to \nu_\tau \pi^-\pi^o\pi^o$;
this is dominated by $\rho\pi$, which originates primarily from the
$a_1(1260)$. Since the $a_1(1260)$ appears clearly here, one
might expect to see the radial excitation $a_1(1700)$ as well.  
This state is interesting as a benchmark for the 2P $n\bar n$ multiplet,
and in view of the reported exotic $\pi_1(1600)$ nearby in mass we must be
especially careful in identifying 2P $q\bar q$ states; nonexotic 
hybrids with $1^{++}$ are predicted 
by the flux-tube model  
to belong to the same 
first hybrid multiplet.
The quark model I=1 2$^3$P$_1$ $q\bar q$ state is
predicted to have a ``fingerprint" decay amplitude; the leading mode,
$\rho\pi$, is predicted to be
dominantly D-wave\cite{Kok,bcps}. 
The S-wave is allowed, but has a node near the physical point due to the
radially excited wavefunction.
This dramatic prediction was apparently confirmed by earlier VES
results\cite{VES_early} and by 
E852 at BNL as reported by Ostrovidov\cite{hybrid_pi1},
but apparently not by recent VES results\cite{Dorofeev} (this 
analysis however is still in progress).

The $a_1(1700)$ may appear as a broadening of the 
$a_1(1260)$ shoulder at high mass in $\tau$ decays; 
the reported $a_1(1260)$ width
of $\sim 600-800$ MeV is much larger than the width observed at E852,
and may indicate the presence of additional states. Separation of the
S- and D- wave $\rho\pi$ amplitudes may be crucial in identifying the
$a_1(1700)$; this was certainly the case for E852.

We note in passing that this state was discussed in $\tau$ hadronic
decays in other sessions by Kravchenko\cite{Kravchenko} (CLEO)
and McNulty\cite{McNulty} (Delphi); both reported improved fits with
an $a_1(1700)$, but due to the dominant $a_1(1260)$ signal the
$a_1(1700)$ could not be identified with confidence. The 
$a_1(1700)$ may also have been seen in $\tau$ decays to 
$\eta\pi\pi\pi$, in the $f_1\pi$ final state. In $f_1\pi$ a clear 
peak does appear at the correct mass\cite{CLEO_f1pi}, 
but there is a concern that the limited $\tau$ phase space
has truncated the resonance shape.
Since the expected branching fractions of the $a_1(1700)$ are 
known,\cite{bcps} and have been measured by E852, it should be 
straightforward to test the strength of the 
possible 
$a_1(1700)$
peak in  $f_1\pi$
at CLEO.

\section{Possible radial excitations reported in other WHS99 sessions}

Many states were reported in talks in other WHS99 sessions which
are plausible candidates for radial excitations. Since these talks will
be reviewed by the appropriate session chairs I will not discuss them
in general in any detail here, but instead simply quote
the
quantum numbers, mass, width, author of the talk and the experiment
(Table~3).
Where these are especially interesting for the subject of radial 
excitations I will discuss the particular state subsequently.

\begin{table}
\centering
\caption{ \it Possible radial excitations reported in other WHS99 sessions. 
}
\vskip 0.1 in
\begin{tabular}{|l|c|c|c|c|} \hline
 $J^{PC}$         &  $M$(MeV) & $\Gamma$(MeV) & mode & contribution \\
\hline
\hline
 I=0:  &  &  &  &   \\
\hline
 $4^{++}$   & $ 2330(30)$  &   $290(70)$ & $\eta\pi^+\pi^-$ & Dorofeev (VES)  \\
 $4^{++}$   & $ 2330(20)$  &   $240(40)$ & $\omega\omega$ & Dorofeev (VES)  \\
\hline
 $2^{++}$   & $ 2310(30)$  &   $230(80)$ & $\eta\pi^+\pi^-$ & Dorofeev (VES)  \\
 $2^{++}$   & $ 2130(35)$  &   $270(50)$ &  $K^+K^-$ & Kirk (WA102) \\
 $2^{++}$   & $ 1980(50)$  &   $450(100)$ &  $\eta\eta$ & Peters  (CBar) \\
 $2^{++}$   & $ 1945(45)$  &   $130(70)$ &  $\eta\eta$ & Kondashov (GAMS) \\
 $2^{++}$   & $ 1940(10)$  &   $150(20)$ & $\omega\omega$ & Dorofeev (VES)  \\
 $2^{++}$   & $ \approx 1645(20)$  
&   $\approx 200(30)$ &  $\pi^o\pi^o, \eta\eta$ & Peters  (CBar) \\
 $2^{++}$   & $ 1645(35)$  &   $230(120)$ &  $\eta\eta$  & Kondashov (GAMS) \\
\hline
 $0^{++}$   & $ 1980(30)$  &   $190(40)$ & $\pi^o\pi^o$, $\eta\eta$ & Kondashov (GAMS)  \\
\hline
\hline
 I=1:  &  &  &  &   \\
\hline
 $3^{--}$   & $ 2300(50)$  &   $240(60)$ & $\eta\pi^+\pi^-$ & Dorofeev (VES)  \\
 $3^{--}$   & $ 2180(40)$  &   $260(50)$ & $\eta\pi^+\pi^-$ & Dorofeev (VES)  \\
\hline
 $2^{++}$   & $ 1752(21)(4)$  &   $150(110)(34)$ &  
$\gamma\gamma\to \pi^+\pi^-\pi^o $ & Braccini (L3) \\
 $2^{++}$   & $ \approx 1670(20)$  
&   $\approx 280(70)$ &  $\eta\pi^o$ & Peters  (CBar) \\
\hline
 $1^{--}$   & $ 2150$ [PDG]   & [PDG]  & $\rho\eta$ & Dorofeev (VES)   \\
 $1^{--}$   & $ 1450$ [PDG]   & [PDG]  & $\rho\eta$ & Dorofeev (VES)   \\
\hline
 $0^{-+}$   & $ 1400(40)$  &   $275(50)$ & $\rho\pi$ & Thoma   (CBar)  \\
\hline
\end{tabular}
\label{table3}
\end{table}

\section{Theoretical aspects of identifying quarkonia and non-quarkonia}

\subsection{Masses}

One might wonder how any of these levels can
be confidently identified as quarkonia, 
since many non-$q\bar q$ states 
with the same quantum numbers are expected.

For the present, masses are our best guide. Fortunately, as one increases
the angular momentum of the $q\bar q$ pair, the multiplet splittings
(spin-spin, spin-orbit, tensor) decrease rapidly. This is because these
are short-distance effects (in the usual OGE and linear confinement
picture), and the centrifical barrier suppresses the short-distance part 
of the $q\bar q$ wavefunction.

For the lowest-mass multiplet of given L$_{q\bar q}$, one can usually identify
the maximum-J state, which is frequently relatively narrow and gives rapidly
varying angular distributions. For example, in the Bugg {\it et al} results
(Tables~1 and 2)
the maximum-J states in the lowest-lying 
L$_{q\bar q}=3, 4$ and 5 multiplets are presumably represented by the
$f_6(2530)$, $\rho_5(2335)$ and $f_4(2025)$. 
These are all well-established PDG states\cite{PDG98}.
Given these levels, we expect the other $n\bar n$ members of these 
1H(2.55), 1G(2.35) and 1F(2.05) multiplets (here rounded to 50 MeV) nearby
in mass.

\begin{table}
\centering
\caption{ \it Suggested multiplets of radially excited 
$n\bar n$ states 
}
\vskip 0.1 in
\begin{tabular}{|l|c|l|} \hline
 nL         &  $M$(GeV) & representative WHS99 candidates \\
\hline
\hline
 2S         & $  1.4  $  &  $\rho(1450),\pi(1300)$    \\   
 3S         & $  1.8  $  &  $\pi(1740)$    \\   
 4S         & $  2.1  $  &  $\rho(2150)$    \\   
\hline
 2P         & $  1.7  $  &  $f_2(1650),a_2(1700),a_1(1700)$    \\   
 3P         & $  2.08 $  &  $f_0(2095),a_1(2100),a_0(2050)$   \\   
 4P         & $  2.34 $  &  $f_0(2335),a_1(2340)$   \\   
\hline
 2D         & $  2.0  $  &  $\omega_3(1950),\eta_2(2040)$ \\   
 3D         & $  2.3  $  &  $\rho_3(2300),\omega_3(2215),\eta_2(2300)$ \\   
\hline
 2F         & $  2.29 $  &  $f_4(2290),f_3(2280),a_4(2280),a_3(2310)$ \\   
\hline
\end{tabular}
\label{table4}
\end{table}
\begin{figure}[t]
\vspace{9.0cm}
\includegraphics{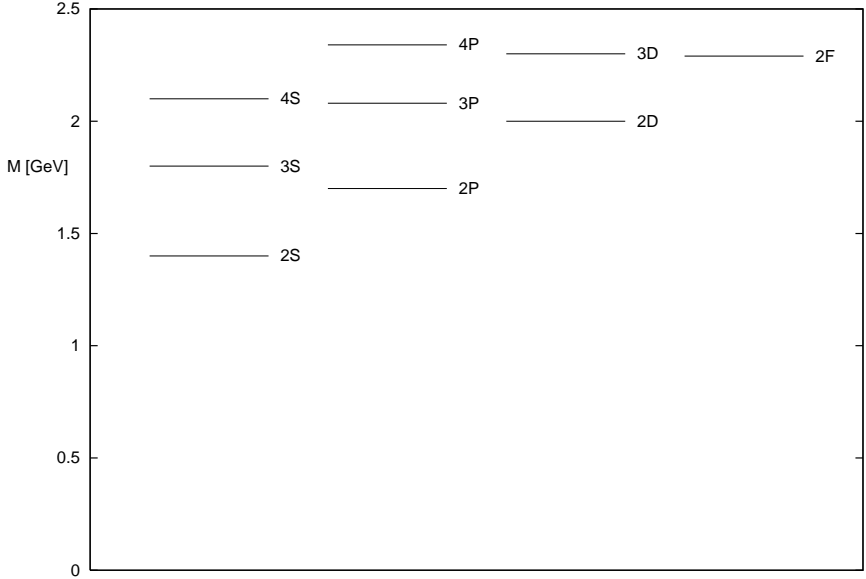}
\caption{\it
Radially-excited $n\bar n$ multiplet levels suggested by recent data
(see Table~4). 
\label{fig1}   }
\end{figure}

Inspection of the tables of states reported in $p\bar p$ (Tables~1 and 2) 
and the higher-mass
states reported in other sessions (Table~3) suggests masses for
some radially-excited $q\bar q$ multiplets, 
which we list in Table~4 and display in Figure~1.
The table includes representative 
candidates discussed in contributions to
WHS99.
Note that these multiplets are rather lighter than expected by
Godfrey and Isgur\cite{God85}, who quoted $n\bar n$ masses for 
radial multiplets
up to
3S, 2P and 2D; their results are 
\begin{equation}
M(3S)|_{\rm GI} = 1.88 {\ \rm GeV \ } (\pi), 
2.00 {\ \rm GeV \ }(\rho),
\label{eq3} 
\end{equation}
\begin{equation}
M(2P)|_{\rm GI} \approx 1.80 {\ \rm GeV},
\label{eq4} 
\end{equation}
\begin{equation}
M(2D)|_{\rm GI} \approx 2.14 {\ \rm GeV}.
\label{eq5} 
\end{equation}
Comparison with Table~4 shows that
experiment is apparently finding these $n\bar n$ radial excitations 
about 0.1-0.2~GeV lower in mass than Godfrey and Isgur
anticipated. For numerical estimates of expected
masses of radially excited levels 
one can of course use ``mass systematics" such as the
radial 
Regge trajectories discussed by Bugg\cite{Bugg} and Peaslee\cite{Peaslee}.
Decay calculations and other matrix elements 
however require explicit meson wavefunctions, which are
usually determined in a
Godfrey-Isgur type model.

\subsection{Strong Decays}
It will be very interesting to see 
if these new, rather low-mass candidates for radially excited $q\bar q$ 
levels can 
still be accommodated in a Coulomb plus linear potential model,
or if there appears to be serious disagreement with this very widely
used description of meson spectroscopy.

In addition to masses and quantum numbers, 
we can expect to have experimental data 
on some relative strong branching fractions. These can be very 
valuable indicators of the nature of a hadron; examples include the
evidence that $\phi=s\bar s$ (a weak $\rho\pi$ mode), 
$f_2(1525)=s\bar s$ (a weak $\pi\pi$ mode), and that $\psi(3097)=c\bar c$
(weak light hadron modes generally). Similarly, discussions of the nature
of the scalar states $f_0(1500)$ and $f_0(1710)$ have centered on
explanations of their strong branching fractions to 
$\pi\pi, K\bar K, \eta\eta$ and $\eta\eta'$. (Here the situation is more
complicated because the theoretical understanding of glueball strong 
decays is not yet well developed.)

For quarkonia there are two commonly used strong decay models,
known as the flux-tube model and the $^3$P$_0$ model. These are rather
similar, in that the assumed decay mechanism in both cases is 
production of a $q\bar q$ pair from the vacuum with $^3$P$_0$ (vacuum)
quantum numbers; this mediates the dominant 
$(Q\bar Q) \to (Q\bar q) (q\bar Q)$ strong decay process. This mechanism
is actually poorly understood and is presumably a nonperturbative effect;
the perturbative OGE pair production amplitude is found in explicit quark model
calculations to be numerically rather weak in most channels\cite{abs}. 

Theorists have carried out detailed decay calculations of higher-mass 
$q\bar q$ states, including all open two-body modes, for $n\bar n$ states
up to 2.1~GeV\cite{bcps}, and for a few specific cases at higher 
mass\cite{Blu}. If these predictions are reasonably accurate they will
be very useful in distinguishing relatively pure $q\bar q$ states
from glueballs, hybrids or molecules.
As specific examples, Tables~5 and 6 
compare the partial widths expected
for quasi-two-body modes of 3$^1$S$_0$ $\pi(1800)$ $q\bar q$ 
and 2$^3$P$_1$ $a_1(1700)$ $q\bar q$ states\cite{bcps} to
expectations for flux-tube hybrids at the same mass.\cite{cp} Evidently
the study of a few characteristic modes, here $\rho\omega$ and
$f_0(1300)\pi$, can
distinguish these assignments. Of course we would prefer to know as
many branching fractions as is experimentally feasible, because the
physics may be more complicated than these simple decay models assume.
In the $\pi(1800)$ case the presence of a strong $\rho\omega$
mode was reported in this meeting by 
VES\cite{Dorofeev},
which argues against this bump being due to a single
flux-tube hybrid. The earlier reports of a large $f_0(1300)\pi$ 
mode argue against a simple $q\bar q$ assignment. 
More than one state may be present, and in the 
most complicated case these 
basis states may be strongly mixed. 

\begin{table}
\centering
\caption{ \it Theoretical two-body partial widths 
(MeV) of a $\pi(1800)$  
}
\vskip 0.1 in
\begin{tabular}{|l|c|c|c|c|c|c|c|} 
\hline
$\pi(1800)$
& $\rho\pi$  
& $\rho\omega$  
& $\rho(1465)\pi$  
&  $f_0(1300)\pi$  
&  $f_2\pi$  
&  $K^*K$  
&  {\rm total}  
\\
\hline
\hline
thy.\cite{bcps} ($q\bar q$)       
&  31 
&  73   
&  53
&   7 
&  28   
&  36
& 228 
\\
thy.\cite{cp} (hybrid)                  
&  30 
&   0   
&  30
& 170 
&   6   
&   5
& $\approx $240 
\\
\hline
\end{tabular}
\label{table5}
\end{table}

\begin{table}
\centering
\caption{ \it Theoretical two-body partial widths 
(MeV) of an $a_1(1700)$  
}
\vskip 0.1 in
\begin{tabular}{|l|c|c|c|c|c|c|c|c|c|} 
\hline
$a_1(1700)$
& $\rho\pi$  
& $\rho\omega$  
&  $b_1\pi$  
& $\rho(1465)\pi$  
&  $f_0(1300)\pi$  
&  $f_1\pi$  
&  $f_2\pi$  
&  $K^*K$  
&  {\rm total}  
\\
\hline
\hline
thy.\cite{bcps} ($q\bar q$)       
&  58 
&  15   
&  41
&  41 
&   2   
&  18   
&  39   
&  33
& 246 
\\
thy.\cite{cp} (hybrid)                  
&  30 
&   0   
& 110  
&   0 
&   6   
&  60   
&  70   
& 20 
& $\approx $300
\\
\hline
\end{tabular}
\label{table6}
\end{table}

One should note that
we still have little information regarding the 
accuracy of our strong decay models for higher-mass states, and more
quantitative information on branching fractions from well-established
$q\bar q$ states is badly needed to allow tests 
of the decay models.
One of the few states discussed at WHS99 for which information 
on relative mode strengths
was reported was in the VES observation of the
$a_4(2040)$. 
Although this is not a radial excitation, 
these data show how detailed
comparisons with theoretical branching fractions 
for radials may be possible in future.
The relative branching fractions 
of the $a_4(2040)$ to $f_2\pi$, $\rho\pi$ and
$\rho\omega$ were reported, which we compare to the predictions of the
$^3$P$_0$ model\cite{bcps} in Table~7. 
(Only modes with theoretical partial widths $> 5$~MeV are tabulated
in Table~7;
for the complete set see Barnes {\it et al.}\cite{bcps}.)
Evidently 
there is good agreement at present accuracy,
which is a nontrivial test of the model
since these modes represent different angular decay amplitudes.

\begin{table}
\centering
\caption{ \it Theoretical and observed 
partial widths
of the $a_4(2040)$
}
\vskip 0.1 in
\begin{tabular}{|l|c|c|c|c|c|c|c|} 
\hline
$a_4(2040)$
& $\eta\pi$  
& $\rho\pi$  
& $\rho\omega$  
&  $b_1\pi$  
&  $f_2\pi$  
&  $KK$  
&  $K^*K^*$  
\\
\hline
\hline
thy.\cite{bcps} ($q\bar q$)       
&  12\ {\rm MeV}   
&  33\ {\rm MeV}   
&  54\ {\rm MeV}   
&  20\ {\rm MeV}     
&  10\ {\rm MeV}     
&  8\ {\rm MeV}     
&  9\ {\rm MeV}     
\\
expt.\cite{Dorofeev}                  
&  -
&  $\equiv 1$   
&  1.5(4)
&  -
&  0.5(2)    
&  -
&  -
\\
\hline
\end{tabular}
\label{table7}
\end{table}

Unfortunately there have been few attempts to measure relative branching 
fractions of higher-mass states, and none were reported for the candidate
radial excitations discussed here. This is an extremely important 
topic for future experimental studies.

\section{The Future of Radials}

Future work on higher-mass $q\bar q$ spectroscopy will hopefully
establish the masses of missing states in the known multiplets
(especially those with masses and quantum numbers expected for 
glueballs and hybrids),
identify the higher-mass states to a mass of at least
$\sim 2.5$~GeV, and 
determine most branching fractions and decay amplitudes 
of a subset of these states
in sufficient detail to be useful to theorists.

Certain $q\bar q$ multiplets are especially interesting because their 
$J^{PC}$ quantum numbers and masses are similar to expectations to
glueballs and hybrids; the $q\bar q$ levels either form a background
and must be identified and eliminated as potential exotica, or they may
mix strongly with the glueball or hybrid states so that the relatively
pure $q\bar q$ level does not exist in nature. Since we cannot say 
{\it a priori} which possibility is correct, it is especially important to 
clarify the experimental spectrum in these mass regions.
Multiplets of special interest for this reason are:

\eject

\vskip 0.2cm
$\bullet{\ 2P}$
\vskip 0.2cm

\begin{table}
\centering
\caption{ \it Theoretical two-body partial widths 
(MeV) of a 2$\, ^3$P$_2$  $f_2(1700)$ $q\bar q$  
}
\vskip 0.1 in
\begin{tabular}{|l|c|c|c|c|c|c|c|c|c|c|c|} 
\hline
$f_2(1700)$
& $\pi\pi$  
& $\eta\eta$  
&  $\eta'\eta$  
& $\rho\rho$  
&  $\omega\omega$  
&  $\pi_{2S}\pi$  
&  $a_1\pi$  
&  $a_2\pi$  
&  $KK$  
&  $K^*K$  
&  {\rm total}  
\\
\hline
\hline
thy.\cite{bcps}       
&  81 
&   4 
&   1 
& 159 
&  56     
&   8    
&  16   
&  43 
&  20  
&  17  
& 405
\\
\hline
\end{tabular}
\label{table8}
\end{table}

\begin{table}
\centering
\caption{ \it Theoretical two-body partial widths 
(MeV) of a 2$\, ^3$P$_2$  $a_2(1700)$ $q\bar q$  
}
\vskip 0.1 in
\begin{tabular}{|l|c|c|c|c|c|c|c|c|c|c|c|c|} 
\hline
$a_2(1700)$
& $\eta\pi$  
& $\eta'\pi$  
&  $\rho\pi$  
& $\rho\omega$  
&  $\eta_{2S}\pi$   
&  $\rho_{2S}\pi$   
&  $b_1\pi$  
&  $f_1\pi$  
&  $f_2\pi$  
&  $KK$  
&  $K^*K$  
&  {\rm total}  
\\
\hline
\hline
thy.\cite{bcps}       
&  23
&  10
& 104
& 109
&   3
&   0
&  28
&   4
&  20
&  20  
&  17  
& 336
\\
\hline
\end{tabular}
\label{table9}
\end{table}

\begin{table}
\centering
\caption{ \it Theoretical decay amplitudes (GeV$^{-1/2}$)
for $f_2(1700)\to\omega\omega$  
}
\vskip 0.1 in
\begin{tabular}{|l|c|c|c|c|} 
\hline
$f_2(1700)$
& $^5$S$_2$ 
& $^1$D$_2$ 
& $^5$D$_2$ 
& $^5$G$_2$ 
\\
\hline
\hline
thy.\cite{bcps} ($q\bar q$)       
&  $+0.34  $
&  $-0.031$
&  $+0.082$
&  $ 0$
\\
\hline
\end{tabular}
\label{table10}
\end{table}

The clarification of the 2P multiplet is 
a high-priority topic;
if the reported exotic $\pi_1(1600)$ is a hybrid,
nonexotic hybrid partners should appear at
a similar mass. We should therefore find an overpopulation 
of states in this mass region,
and perhaps anomalous decay branching fractions if there is strong 
quarkonium-hybrid mixing.
The flux-tube model expects 
exotic partners with 
$J^{PC}=0^{+-}$ and $2^{+-}$, and nonexotics with
$J^{PC}=0^{-+},1^{+-},1^{--},1^{++}$ and $2^{-+}$, all at about the same
mass as the $1^{-+}$. 
Thus we expect an apparent duplication of the
2$^1$P$_1$ and 2$^3$P$_1$ 
2P quarkonium levels by hybrids at similar masses.

Expected branching fractions for flux-tube hybrids have been
calculated in detail by Close and Page\cite{cp}, and these modes should be
studied for evidence of both hybrid and quarkonium 
parent resonances.

Studies of 2P candidates that are {\it not} expected to have nearby 
hybrids are important as calibration studies for the decay models. 
The 2$^3$P$_2$ states 
$a_2(1670-1750)$ and $f_2(1645)$ reported at this meeting may be most 
straightforward to study, since neither hybrids nor glueballs are 
expected in $2^{++}$ at this mass. The theoretical partial widths
for 2$^3$P$_2$ 
$f_2(1700)$ 
and
$a_2(1700)$ quarkonia
are given in Tables~8 and 9. Although the $f_2(1700)$ state
is reported in $\pi^o\pi^o$ and $\eta\eta$ and
the Crystal Barrel reports the $a_2(1700)$ in $\eta\pi^o$, 
these modes are not
predicted to be dominant; for $f_2(1700)$ 
$\rho\rho$ is expected to be 
largest, and an experimentally clean $\omega\omega$ mode with
$1/3$ the $\rho\rho$ strength is also predicted.
Similarly the $a_2(1700)$ should have a large $\rho\omega$ mode.
These $VV$ modes are quite interesting in that there are several 
subamplitudes, and the predicted amplitude ratios are nontrivial.
As an example, the numerical $^3$P$_0$ decay 
amplitudes for $f_2(1700)\to\omega\omega$ 
are given in Table~10. If the S-wave and D-wave $\omega\omega$
amplitudes from this state could be separated and compared, we would have
a very sensitive test of the decay model, and if there is
agreement we could apply the same model with
more confidence to the decays of other candidate 
high-mass $q\bar q$ states.

As a final observation, the report 
by Braccini (L3 Collaboration)\cite{Braccini}
of a 2P candidate $a_2(1750)$
in $\gamma\gamma$ collisions 
is especially interesting because this is the first radial excitation
to be reported in $\gamma\gamma$. Theoretically, radially excited 
$q\bar q$ states should appear with little suppression in $\gamma\gamma$
collisions,\cite{twophot} but to date only this radial candidate has been 
reported. There may actually be a problem here, as the mass reported 
for an
excited $a_2$ state by Crystal Barrel,\cite{Peters} 
$\approx 1670(20)$~MeV, does not appear consistent with the L3 mass.

\vskip 0.2cm
$\bullet{\ 3P \ {\rm and} \ 4P}$
\vskip 0.2cm

These multiplets, which experimentally lie at about 2.08~GeV and 2.34~GeV
(for $n\bar n$; see Table~4), 
will be of interest because of the presence of the 
lightest tensor glueball. Of course there will be hybrids in this
region as well, so we can expect a complicated spectrum of overlapping
resonances.
At present there is no theoretical guidance regarding decay modes
of these $q\bar q$ states; 
such a study would be difficult to motivate without evidence
(for example from the 2P multiplet) that the decay models will
give useful results
for these high radial excitations.

\vskip 0.2cm
$\bullet{\ 2S \ {\rm and} \ 3S}$
\vskip 0.2cm

The 2S and 3S multiplets are also interesting due to their
proximity to the reported hybrid candidates $\pi_1(1405)$ and
$\pi_1(1600)$; both the bag model and flux-tube model
predict that the lightest $1^{-+}$ hybrid has
$0^{-+}$ and $1^{--}$ partners nearby in mass.
These should be observable as an overpopulation of states 
in the 
L$_{q\bar q}=0$ $q\bar q$ sectors.

The 2S multiplet has historically been problematic because there
are broad overlapping states in the $1^{--}$ sector; at least
two states near 1.45 and 1.7~GeV
are needed to explain the data (notably 
$e^+e^- \to \pi\pi $ and $\omega\pi$). 
The observation of the $\rho(1450)$
in $\tau$ decays at CLEO was reported here by 
Kravchenko\cite{Kravchenko}, who also noted the absence of 
a $K^*(1410)$ signal in $K\pi$. 
The topic of vector meson spectroscopy in this mass region
was recently reviewed by Donnachie and 
Kalashnikova,\cite{DK} who concluded that an additional vector
was required 
in both I=0 and I=1 channels
to fit the data. 
In I=1, the 
weakness of
$e^+e^-\to \pi^+\pi^-\pi^o\pi^o$ relative to
$e^+e^-\to \pi^+\pi^-\pi^+\pi^-$ cannot be
explained by the expected $\rho(1700)$
decays alone. 
This additional state may be the $1^{--}$ vector
hybrid, which is expected in both bag and flux-tube models.
The recent work on $4\pi$ decays of I=1 vectors
by the Crystal Barrel Collaboration\cite{CB_4pi} has 
shown that the 
$a_1\pi$ and $h_1\pi$ 
modes come from the 1.7~GeV region rather than
1.45~GeV, as expected if
the 1.7~GeV state is the 1D $q\bar q$ state and the 1.45~GeV 
is 2S, which is the usual quark model assignment. 
Clearly a more detailed study of the different $4\pi$ charge
states would be enlightening.

In the 2$^1$S$_0$ states,
the decay of the $\pi(1300)$ is observed by Crystal
Barrel\cite{Thoma} to be dominantly $\rho\pi$ rather than $(\pi\pi)_S\pi$,
which is in agreement with the expectations of the $^3$P$_0$
model. Of course the $\pi(1800)$ state has attracted considerable 
attention as a possible hybrid, largely because of the $f_0(1300)\pi$
decay mode (see Table~5). The VES report of a large $\rho\omega$ 
mode\cite{Dorofeev} at about 1.74~GeV was attributed 
by
Dorofeev
to an overpopulation
of $\pi(1740-1800)$ states, and we should presumably
identify the 3S quark model state identified
with the large $\rho\omega$ signal. This can be tested by studies of
$f_2\pi$ and $K^*K$; a 3$^1$S$_0$ $q\bar q$ $\pi(1800)$ is
predicted to have branching fractions into these modes
about half 
as large as its $\rho\omega$ branching fraction (see Table~5).

\section{Acknowledgements}
It is a pleasure to acknowledge the organizers of WHS99
for their kind invitation to arrange 
a session on radial excitations. I would also 
like to thank them for the opportunity
to participate
in discussions of the physics of hadrons with our colleagues.
I am also grateful to
C.Amsler,
R.Baker,
N.Black,
D.V.Bugg,
S.-U.Chung,
F.E.Close,
V.Dorofeev,
R.Galik,
T.Handler,
N.Isgur,
L.Montanet,
A.Ostrovidov,
E.S.Swanson
and
H.J.Willutski
for discussions of the particular aspects of meson spectroscopy
presented here. Research at the Oak Ridge National Laboratory is
supported by the U.S. Department of Energy under contract
DE-AC05-96OR22464 with Lockheed Martin Energy Research Corp.

\end{document}